\documentstyle{article}

\input{epsf.tex}

\newcommand {\tsub}[1]{_{\mbox{\protect\scriptsize #1}}}

\newcommand {\Ref}[1]{(\ref{#1})}
\newcommand {\matrixelm}[3]{
\left\langle#1\right|#2\left|#3\right\rangle}
\newcommand {\innerm}[2]{
\left\langle#1\left|#2\right.\right\rangle}
\newcommand {\dirac}[1]{$\left|#1\right\rangle$}

\newcommand {\diracm}[1]{\left|#1\right\rangle}

\newcommand {\Stretch}[1]{
\renewcommand{\baselinestretch}{#1 }\large\normalsize}

\newcommand {\bm}[1]{{\mbox{\boldmath $#1$}}}

\newcommand {\citex}[1]{\protect\cite{#1}} 

\begin{document}

\Stretch{3}

\begin{center}

{\Large \bf Molecular Orbital Models of Benzene, Biphenyl and the
Oligophenylenes }

\bigskip

\Stretch{1.5} {\large Robert J.\ Bursill$^1$, William Barford$^2$
and Helen Daly$^2$}

$^1$School of Physics, University of New South Wales, Sydney, NSW 2052, Australia.\\
Email: ph1rb@newt.phys.unsw.edu.au

$^2$Department of Physics, The University of Sheffield, Sheffield, S3 7RH, U.\ K.\\
Email: w.barford@sheffield.ac.uk

\end{center} \Stretch{1}

\begin{abstract}

A two state (2-MO) model for the low-lying long axis-polarised excitations of poly({\em 
p}-phenylene) oligomers and polymers is developed. First we derive such a model from 
the underlying Pariser-Parr-Pople (P-P-P) model of $\pi$-conjugated systems. The two 
states retained per unit cell are the Wannier functions associated with the valence and 
conduction bands. By a comparison of the predictions of this model to a four state model 
(which includes the non-bonding states) and a full P-P-P model calculation on benzene 
and biphenyl, it is shown quantitatively how the 2-MO model fails to predict the correct 
excitation energies. The 2-MO model is then solved for oligophenylenes of up to 15 
repeat units using the density matrix renormalisation group (DMRG) method. It is shown 
that the predicted lowest lying, dipole allowed excitation is ca.\ 1 eV higher than the 
experimental result. The failure of the 2-MO model is a consequence of the fact that the 
original HOMO and LUMO single particle basis does not provide an adequate 
representation for the many body processes of the electronic system. 

\end{abstract}

\section{Introduction}
\label{introduction}

Interest in the low-lying excitations of the phenyl based semiconductors, in particular 
poly({\em p}-phenylene) and poly({\em p}-phenylene vinylene), arises from the 
observation of their electroluminescence \cite{bradley1}, \cite{klemenc} and the 
possibility of various optical and nonlinear devices. From a theoretical point of view one 
would like to understand how the excitations of the phenyl based semiconductors are 
derived from the parent excitations of benzene, how these evolve as a function of 
oligomer length, and how they participate in non-linear optical processes. 

There is now a substantial body of experimental results on the photo- and electro- 
luminescent properties of poly({\em p}-phenylene vinylene) 
\cite{bradley2,bradley3,bradley4,bradley5,heeger,leng,mazumdar,kersting}, and this has 
generated commicant theoretical interest \cite{mazumdar,bredas1,ssh,rice,shimoi}. Fewer 
experimental results exist for poly({\em p}-phenylene) 
\cite{tieke,ambrosch,shacklette,klemenc,tasch}, however, largely resulting from the 
difficulties in obtaining well characterised material. There have been a number of 
theoretical calculations on poly({\em p}-phenylene). Br\'{e}das has used the VEH 
pseudopotential technique \cite{bredas2} and Ambrosch-Draxl et al.\ have performed 
density functional calculations using LAPW and pseudopotentials \cite{ambrosch}. Rice 
et al.\ \cite{rice} have developed a phenomenological, microscopic model based on the 
molecular excitations of benzene. The absorption bands are calculated using an 
approximate Kubo formalism.

In this paper we will restrict our attention to poly({\em p}-phenylene), and develop in full 
detail the model and computational techniques introduced in a recent letter \cite{bb1}, 
\cite{bb2}. Our goal is to construct a model of poly({\em p}-phenylene) based on the 
underlying Pariser-Parr-Pople (P-P-P) model of $\pi$-conjugated electron systems 
\cite{pariser}. The P-P-P model has long been used to describe the low-lying excitations 
of $\pi$-conjugated systems, giving reasonable results. However, an improved 
parameterisation of this model is possible, and that was achieved in a previous paper 
\cite{bb3}. We will use this optimised parameterisation in the current paper. We will 
show, at 
the very least, that to achieve our goal of a full description of 
poly({\em p}-phenylene) a four molecular orbital (4-MO) model (as 
described below) is required. However, in this paper our ambition will 
be more limited, and we will primarily be concerned with a description 
of the long axis-polarised excitations. 
One of the aims of this paper is to show 
how well the 2-MO model, whose parameters are directly obtained from 
the underlying P-P-P model, explains the physics of the oligophenylenes. 
This will be done by comparing the predictions of the 4-MO and 2-MO models 
to exact P-P-P calculations of benzene and biphenyl. By systematically 
reducing the size of the Hilbert space we will show how the 
discrepencies between the full and reduced models arise. Next, equipped 
with the knowledge of how well the 2-MO model predicts the biphenyl 
excitations, we solve oligophenylenes of up to 15 repeat units using 
the density matrix renormalisation group (DMRG) method. This technique 
is ideally suited to solving lattice quantum Hamiltonians with open 
boundary conditions. We find that the theoretical predictions of
the exciton energies are ca.\ 1 eV higher than the experimental results.
We argue that failure of the 2-MO model is a consequence of the fact 
that the original HOMO and LUMO single particle basis does not provide 
an adequate representation for the many body processes of the electronic
 system.

\begin{figure}[htbp]
\caption{
Molecular orbitals of phenylene, the repeat unit of poly({\em p}-phenylene). The 
amplitude of the wavefunction is indicated by shading. The dashed lines are the nodes of 
the wavefunction.
}
\centerline{\epsfxsize=15.5cm \epsfbox{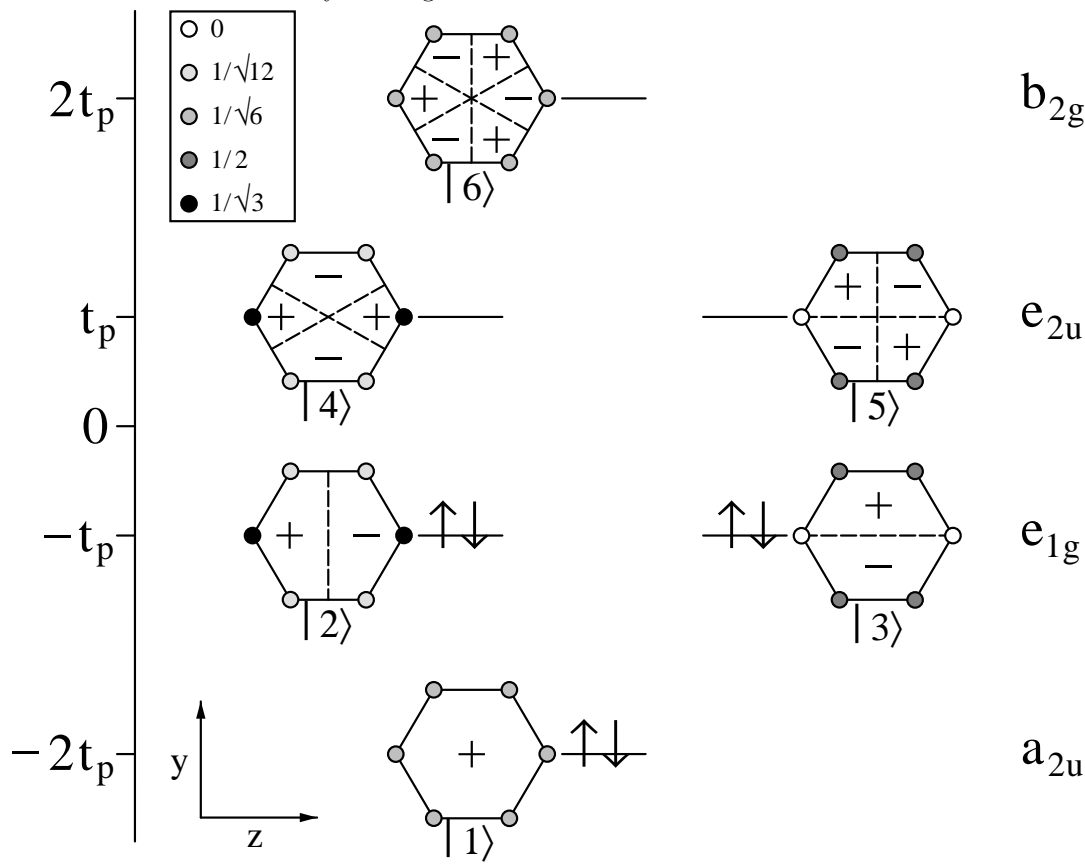}}
\end{figure}

The molecular orbital approach to the study of conjugated polymers is
not new.  Recently it has been used by Soos {\it et al.} \cite{soos92}
on their work on the phenyl semiconductors,
while Chandross  {\it et al.} employed a similar approach in their
work on conjugated polymers \cite{chandross97}.

The plan of this paper is as follows. First, the 4-MO and 2-MO models of 
benzene and oligophenylenes (including biphenyl) will be introduced. By 
making direct comparison between the full 6-MO calculation of benzene 
and biphenyl we will show the successes and limitations of these reduced 
models. Next, we solve the 2-MO model to describe the low energy physics 
of the oligophenylenes. Finally, section~\ref{conclusions} concludes.

\section{The Molecular Orbital Approach}
\label{MO}

\subsection{The Four-Molecular Orbital (4-MO) Model} 

The essential repeat unit of oligophenylenes is the phenylene repeat unit. Its one-electron 
atomic orbital basis may be more conveniently represented as a one-electron molecular 
orbital basis, which is the set of eigenstates of the phenylene kinetic energy operator. 
These are illustrated in Fig.\ 1 as eigenstates of the $x$-$y$ and $x$-$z$ plane reflection 
operators (i.e.\ $\hat{\sigma}(xy)$ and $\hat{\sigma}(xz)$). Solving the P-P-P model 
within the full 6-MO basis is of course equivalent to solving within the full atomic orbital 
basis. However, keeping such a large number of states within an exact calculation soon 
becomes prohibitively expensive in memory and cpu time resources. The largest 
oligophenylene which can be solved exactly is biphenyl. Even with sophisticated Hilbert 
space truncation procedures, such as the DMRG method, there are technical reasons why 
it is difficult to retain the full 6-MO, or even the 4-MO, basis.

In addition to the technical difficulties of retaining the full 6-MO basis, one might 
suppose that the low-lying excitations arise between the MO states closest to the Fermi 
energy, and that these alone are sufficient to describe the low energy physics of 
oligophenylenes. In this approach, the MOs furthest from the Fermi energy remain 
frozen. In benzene, for example, the MOs which are retained are the $e_u$ and $e_g$ 
states, while the $a_u$ and $b_g$ states are frozen. In biphenyl, extended MOs are 
formed, and, by Fourier transforming the extended orbitals, we construct localised 
(Wannier) orbitals. The eight Wannier orbitals which are retained are those associated 
with the eight biphenyl MOs closest to the Fermi energy.

\subsection{The Two-Molecular Orbital (2-MO) Model}

The 2-MO model reduces the Hibert space yet further by retaining only a pair of states 
per repeat unit. In benzene, four calculations are performed, corresponding
to the various pairings of the bonding ($|2\rangle$ and $|4\rangle$)
and non-bonding ($|3\rangle$ and $|5\rangle$) orbitals. In biphenyl it is
either the Wannier orbitals associated with the bonding biphenyl 
MOs or the non-bonding orbitals. Likewise, in long oligophenylenes,
the extended MOs 
form bands, as shown in Fig.\ 2 for an infinite chain with periodic boundary conditions. 
By Fourier transforming the Bloch states associated with the valence and conduction 
bands we obtain the pair of relevant Wannier orbitals required to describe the low energy 
physics, as shown in Appendix A \cite{footnote}. The reason why this might appear
to be a reasonable assumption for the consideration of the $1^1B_{1u}$ 
exciton of oligophenylenes will be discussed in more detail shortly. 

\begin{figure}[htbp]
\caption{
Non-interacting band structure of poly({\em p}-phenylene).
}
\centerline{\epsfxsize=15.5cm \epsfbox{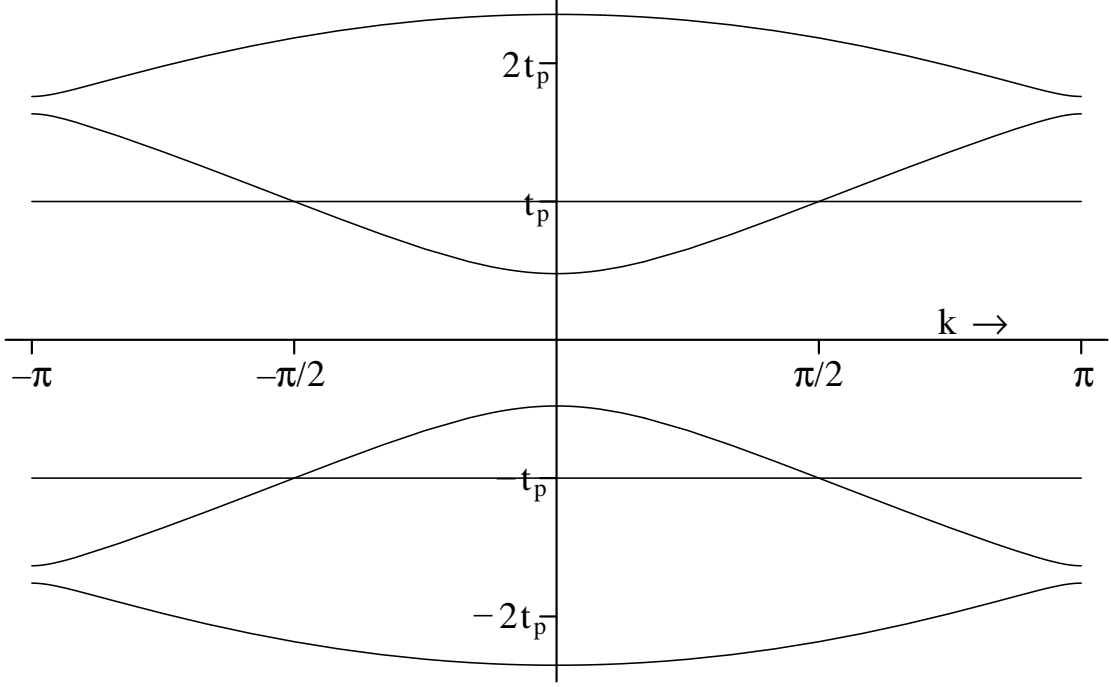}}
\end{figure}

\section{Benzene}
\label{benzene}

In this section we compare the predictions of the 4-MO and 2-MO models to the full 
(6-MO) P-P-P model calculation of benzene.

The P-P-P Hamiltonian is written as
\begin{equation}
H = - \sum_{<ij>\,\sigma} t_{ij} \left[ c_{i\sigma}^{\dagger} c_{j\sigma} + \rm{ h.c.} 
\right] + U \sum_{i} \left( n_{i\uparrow} - \frac{1}{2} \right) \left( n_{i\downarrow} - 
\frac{1}{2} \right) + \frac{1}{2} \sum_{i\neq j} V_{ij} (n_i - 1)(n_j - 1),
\label{ppp_hamiltonian}
\end{equation}
where $c_{i\sigma}^{\dagger}$ creates a $\pi$ electron with spin $\sigma$ on carbon 
site $i$, $n_{i\sigma} = c_{i\sigma}^{\dagger}c_{i\sigma}$, $n_{i} = n_{i\uparrow} + 
n_{i\downarrow}$ and $<>$ represents nearest neighbours.

We use the Ohno parameterisation for the Coulomb interaction \cite{ohno},
\begin{equation}
V_{ij} = \frac{U}{ (1 + \alpha r_{ij}^2)^{1/2} },
\end{equation}
where $\alpha=(U/14.397)^2$, 
thus ensuring that $V_{ij} \rightarrow e^2/(4\pi\epsilon_0 
r_{ij})$ as $r_{ij} \rightarrow \infty$, and $r_{ij}$ is the inter-atomic distance in \AA. 
The C-C bond length is taken as 1.40 \AA. The optimal parameterisation, which was 
derived in \cite{bb3}, is $U=10.06$ eV and the phenyl bond transfer integral, 
$t\tsub{p}=2.539$ eV.

The MO representation diagonalises the kinetic energy operator at the expense of 
introducing off-diagonal, two-electron terms into the interactions. In this representation 
the Hamiltonian reads
\begin{equation}
H=\sum_{\alpha=1}^6 \epsilon_{\alpha} (n_{\alpha} -1) + \frac{1}{4} 
\sum_{\alpha\beta\gamma\delta\sigma \sigma^{\prime}} 
V_{\alpha\beta\gamma\delta}(a^{\dagger}_{\alpha\sigma} 
a^{\dagger}_{\gamma\sigma^{\prime}} a_{\delta\sigma^{\prime}}a_{\beta\sigma}
+a_{\bar{\alpha}\bar{\sigma}} a_{\bar{\gamma}\bar{\sigma^{\prime}}} 
a^{\dagger}_{\bar{\delta}\bar{\sigma^{\prime}}}
a^\dagger_{\bar{\beta}\bar{\sigma}}),
\end{equation}
where
\begin{equation}
V_{\alpha \beta \gamma \delta }= \int \int d^3{\bf r}_1 d^3{\bf r}_2\, 
\psi^{\ast}_{\alpha}({\bf r}_1) \psi_{\beta}({\bf r}_1) \frac{e^2}{4\pi\epsilon_0|{\bf 
r}_1-{\bf r}_2|} \psi^{\ast}_{\gamma}({\bf r}_2) \psi_{\delta}({\bf r}_2),
\end{equation}
$a_{\alpha\sigma}^{\dagger}$ creates an electron of spin $\sigma$ in the MO 
\dirac{\alpha} and $\epsilon_{\alpha}$ are the MO eigenvalues.  Eqn.\ (3) is invariant 
under the particle-hole transformation $a^{\dagger}_{\alpha\sigma}
\leftrightarrow {\rm sgn}(\sigma) 
a_{\bar{\alpha}\bar{\sigma}}$ where the MOs transform as \dirac{\bar{1}} = 
\dirac{6},  \dirac{\bar{2}} = \dirac{4} and  \dirac{\bar{3}} = \dirac{5}.

Inserting
\begin{equation}
\psi_{\alpha}({\bf r}_1)=\sum_i f^{\alpha}_i \phi_i({\bf r}_1),
\end{equation}
where $ \phi_i({\bf r}_1)$ are the orthogonalised atomic orbitals, and using
\begin{equation}
\int \int d^3{\bf r}_1 d^3{\bf r}_2\, \phi^{\ast}_i ({\bf r}_1)\phi_{i^{\prime}}({\bf r}_1) 
\frac{e^2}{4\pi\epsilon_0|{\bf r}_1-{\bf r}_2|} \phi^{\ast}_j({\bf 
r}_2)\phi_{j^{\prime}}({\bf r}_2) =V_{ij}\delta_{i i^{\prime}}\delta_{j j^{\prime}}
\end{equation}
gives
\begin{equation}
V_{\alpha \beta \gamma \delta }= \sum_{ij} f^{\alpha}_i f^{\beta}_i V_{ij} 
f^{\gamma}_j f^{\delta}_j .
\end{equation}
Table~\ref{2-e} lists the two-electron integrals used in the benzene calculation. All the 
two-electron integrals are retained in the 6-MO calculation to reproduce the atomic 
orbital basis calculation of \cite{bb3}.

\begin{table}[htbp]

\centering

\begin{tabular}{|c|c|c|c|c|}
\hline 2-e parameter & Name & Benzene & Biphenyl & Oligophenylenes \\
\hline $V_{\alpha\alpha\alpha\alpha}$ & `Onsite' Coulomb & Yes & Yes & Yes \\
\hline $V_{\alpha\alpha\beta\beta}$ & Direct Coulomb & Yes & Yes & Yes \\
\hline $V_{\alpha\beta\beta\alpha}$ & Exchange & Yes & Yes & Yes \\
\hline $V_{\alpha\beta\alpha\beta}$ & Pair hop & Yes & Yes & Yes \\
\hline $V_{\alpha\alpha\alpha\beta}$ & Effective hopping & Yes & Yes & No \\
\hline $V_{\alpha\alpha\beta\gamma}$ & Effective hopping & Yes & Yes & No \\
\hline $V_{\alpha\beta\alpha\gamma}$ & 3-centre & Yes & Yes & No \\
\hline $V_{\alpha\beta\gamma\alpha}$ & 3-centre & Yes & Yes & No \\
\hline $V_{\alpha\beta\gamma\delta}$ & 4-centre & Yes & Yes & No \\
\hline
\end{tabular}

\caption{The two-electron parameters used in the MO Hamiltonian, Eq.\ (3). }
\label{2-e}
\end{table}

A reduced symmetry of the benzene Hamiltonian is invariance under reflection in the 
$x$-$y$ and $x$-$z$ planes. The MO basis, Fig.\ 1, possesses this symmetry and partly 
diagonalises the many body Hamiltonian (3). In particular, the states arising from 
excitations between (predominately) $\diracm{2}\rightarrow\diracm{4}$ and 
$\diracm{3}\rightarrow\diracm{5}$ mix and are odd under $\hat{\sigma}(xy)$ and even 
under $\hat{\sigma}(xz)$ reflection. These become the $1B_{1u}$ and $1E_{1u}(z)$ 
states. Similarly, the states arising from excitations between (predominately) 
$\diracm{2}\rightarrow\diracm{5}$ and $\diracm{3}\rightarrow\diracm{4}$ mix and are 
odd under $\hat{\sigma}(xz)$ and even under $\hat{\sigma}(xy)$ reflection, and these 
become the $1B_{2u}$ and $1E_{1u}(y)$ states.

\begin{table}[htbp]

\centering

\begin{tabular}{|c|c|c|c|c|c|c|}
\hline State & $\sigma(xy)$ & $\sigma(xz)$ & 6-MO & 4-MO & 2-MO & Experiment \\
\hline $1^1B^+_{2u}$ & $+$ & $-$ & 4.75 & 5.55 & 6.34 & 4.90 \\
\hline $1^1B^-_{1u}$ & $-$ & $+$ & 5.47 & 5.71 & 6.43 & 6.20 \\
\hline $1^1E^-_{1u}(z)$ & $-$ & $+$ & 6.99 & 7.34 & 6.43 & 6.94 \\
\hline $1^1E^-_{1u}(y)$ & $+$ & $-$ & 6.99 & 7.34 & 6.34 & 6.94 \\
\hline
\hline $1^3B^+_{1u}$ & $-$ & $+$ & 4.13 & 4.68 & 4.73 & 3.94 \\
\hline $1^3E^+_{1u}(z)$ & $-$ & $+$ & 4.76 & 5.24 & 4.73 & 4.76 \\
\hline $1^3E^+_{1u}(y)$ & $+$ & $-$ & 4.76 & 5.24 & 5.08 & 4.76 \\
\hline $1^3B^-_{2u}$ & $+$ & $-$ & 5.60 & 5.55 & 5.08 & 5.60 \\
\hline
\end{tabular}

\caption{Full P-P-P (6-MO), 4-MO and 2-MO model calculations of the vertical, 
low-lying excitations of benzene in eV. Also listed are the experimental vertical, low-lying 
transitions from \citex{CASPT2_benzene}. }
\label{benzene.table}
\end{table}

Table~\ref{benzene.table} lists the results of the 6-MO (full) and the 4-MO calculation. 
The full 6-MO basis, using the optimised parameters derived in \cite{bb3}, results in an 
average error of 2.75\%. The ordering of the states is consistent with experiment, except 
for the $1^1B_{2u}$ state, which lies just above the $1^3E_{1u}$ state experimentally, 
while theoretically the ordering is reversed. However, the energy of the $1^1B_{1u}$ 
state is over $0.7$ eV too low \cite{footnote2}. 

The discrepencies between the full calculation and the 4-MO results are reasonable 
(particularly for the dominant singlet excitations, $1^1E_{1u}$), being of the order of a 
few tenths of an eV. The 4-MO calculation, however, does predict that the singlet and 
triplet $1B_{2u}$ states are degenerate, whereas experimentally the singlet lies below the 
triplet.

The results of the 2-MO calculations are also shown. The 
$\diracm{2}\rightarrow\diracm{4}$ and $\diracm{3}\rightarrow\diracm{5}$ excitations 
decouple and are degenerate, and likewise with the $\diracm{2}\rightarrow\diracm{5}$ 
and $\diracm{3}\rightarrow\diracm{4}$ excitations. As expected, there are substantial 
deviations with these results from the full calculation. This results from the absence of 
mixing between the states. However, for oligophenylenes, the bonding HOMO and 
LUMO states will begin to form bands, thus lifting the degeneracy between the bonding 
and non-bonding orbitals, and hence the mixing will reduce. Thus, for oligomers the
2-MO model may result in smaller discrepancies. To investigate this assumption, we now 
turn to a discussion of biphenyl. 

\section{Biphenyl}
\label{biphenyl}

The accuracy of the 4-MO and 2-MO models as applied to biphenyl will give some 
indication of their reliability for longer oligophenylenes. As in benzene, the 4-MO model 
uses four states per repeat unit. However, rather than use the primitive benzene MOs, we 
construct Wannier MOs obtained by Fourier transforming the biphenyl molecular 
orbitals. A full description of this procedure is left to Appendix A, where Wannier MOs 
are derived for a system with periodic boundary conditions.

\begin{figure}[htbp]
\caption{
Chemical structure of biphenyl showing the bond lengths used in our calculation.
}
\centerline{\epsfxsize=15.5cm \epsfbox{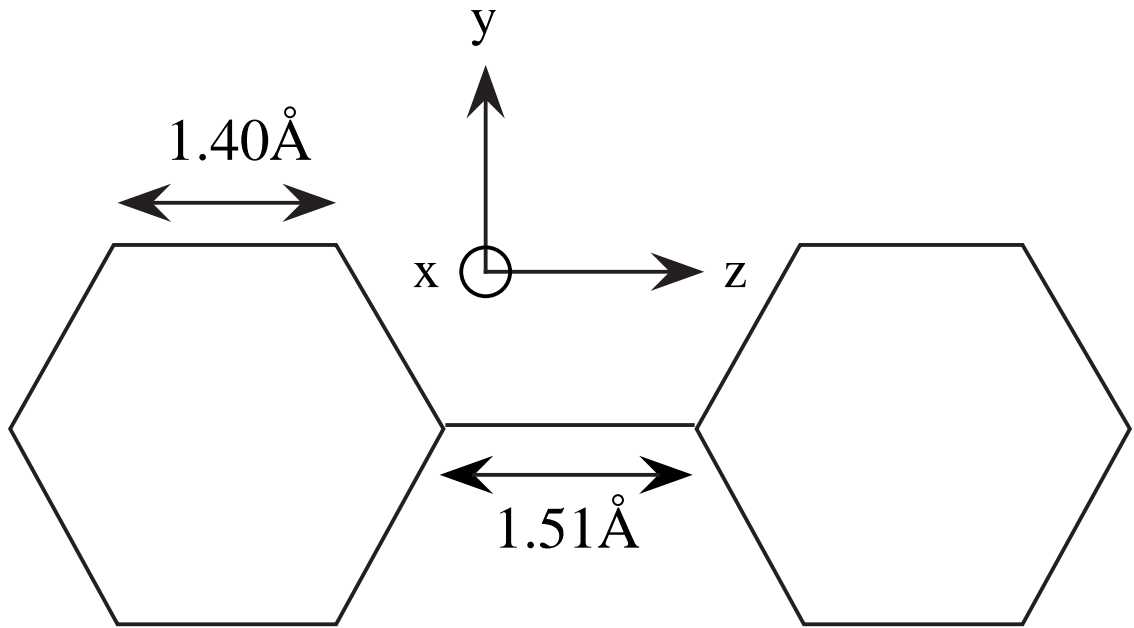}}
\end{figure}

Biphenyl belongs to the $D_{2h}$ symmetry group. We adopt the convention that the 
$z$-axis is the long axis and the $y$-axis is the short axis, as shown in Fig.\ 3, along with 
the bond lengths used in our calculation. The optimised parameterisation for $U$ and 
$t\tsub{p}$ are those derived in \cite{bb3} and used in section 3. The single bond 
hybridisation integral, $t\tsub{s}$, is also determined empirically by fitting the 
theoretical calculation to the experimental $1^1B_{1u}$ exciton at $4.80$ eV in biphenyl 
crystals. This gives $t\tsub{s} = 2.22$ eV.

\begin{table}[htbp]

\centering

\begin{tabular}{|c|c|c|c|c|c|c|}
\hline State & $\sigma(xy)$ & $\sigma(xz)$ & 6-MO & 4-MO & 2-MO & Experiment \\
\hline $1^1B^+_{2u}$ & $+$ & $-$ & 4.55 & 5.35 & NA & 4.20 (0-0)--4.49\\
\hline $1^1B^+_{3g}$ & $-$ & $-$ & 4.58 & 5.40 & NA & 4.11 (0-0) \\
\hline $1^1B^-_{1u}$ & $-$ & $+$ & 4.80 & 5.25 & 5.46 & 4.80 \\
\hline $2^1A^-_{g}$ & $+$ & $+$ & 5.57 & 5.85 & 6.80 & 4.71--5.02\\
\hline $2^1B^-_{1u}$ & $-$ & $+$ & 6.22 & 6.59 & 6.43 & 6.14, 6.16 \\
\hline $2^1B^-_{3g}$ & $-$ & $-$ & 6.28 & 6.80 & NA & --- \\
\hline $3^1A^+_{g}$ & $+$ & $+$ & 6.30 & 7.19 & 7.14 & ca.\ 6.0 (max) \\
\hline $2^1B^-_{2u}$ & $-$ & $+$ & 6.66 & 7.17 & NA & 5.85, 5.96 \\
\hline
\hline $1^3B^+_{1u}$ & $-$ & $+$ & 3.63 & 4.29 & 4.42 & ca.\ 3.5 (max) \\
\hline $1^3A^+_{g}$ & $+$ & $+$ & 4.25 & 4.84 & 5.13 & --- \\
\hline $2^3B^+_{1u}$ & $-$ & $+$ & 4.56 & 5.06 & 4.73 & --- \\
\hline $1^3B^+_{2u}$ & $+$ & $-$ & 4.56 & 5.10 & NA & 3.93 (0-0) \\
\hline $1^3B^+_{3g}$ & $-$ & $-$ & 4.56 & 5.10 & NA & --- \\
\hline $2^3A^+_{g}$ & $+$ & $+$ & 4.80 & 5.35 & 4.73 & --- \\
\hline $2^3B^-_{2u}$ & $+$ & $-$ & 5.32 & 5.42 & NA & --- \\
\hline $2^3B^-_{3g}$ & $-$ & $-$ & 5.37 & 5.47 & NA & --- \\
\hline
\end{tabular}

\caption{Full P-P-P (6-MO), 4-MO and 2-MO model calculations of the low-lying 
excitations of biphenyl in eV. The experimental results are described in \citex{bb3}.}
\label{biphenyl.table}
\end{table}

Table \ref{biphenyl.table} shows the full P-P-P biphenyl calculation and the corresponding experimental 
results. A full comparison of theory and experiment is presented in ref.\ \cite{bb3}, so 
here we merely summarise the results. The P-P-P calculation is very successful in its 
predictions of the long axis-polarised singlet and triplet states. It is less succesful, 
however, in its treatment of the short axis-polarised states. For example, it predicts that 
the $2^1B_{2u}$ state is higher than the $2^1B_{1u}$ state and that the $1^1B_{3g}$ 
state lies higher than the $1^1B_{2u}$ state, both of which are in contradiction to 
experiment. In \cite{bb3} we argued that this discrepency is a consequence of the neglect 
of next nearest neighbour hopping in the P-P-P model.

Now let us compare the predictions of the 4-MO and 2-MO model calculations to the full 
calculation. As shown in table~\ref{biphenyl.table} the most obvious discrepency 
between the 4-MO and full calculations is that the former predicts that the
$1^1B^-_{1u}$ state lies below the $1^1B^+_{3g}$ and $1^1B^+_{2u}$ states. In addition, all 
the energies are too high: the $1^1B^-_{1u}$ state is predicted at 0.45 eV (i.e.\ 9\%) 
higher than the full calculation. The ordering of the states in the triplet sector agrees with 
the full calculation, but again they ca.\ 0.5 eV too high. This poor agreement between the 
6-MO and 4-MO calculations indicates the importance of electronic correlations between 
the eight active orbitals (i.e. the HOMO and LUMO) and the four frozen orbitals (i.e. 
those furthest from the Fermi energy). The neglect of the strong mixing of these frozen 
states in the 4-MO calculation results in both quantitative and qualitative discrepencies. 

Table~\ref{biphenyl.table} also shows the results of the 2-MO calculation for the long 
axis-polarised states. It is noteworthy that the discrepencies between the 2-MO and 
4-MO calculations are smaller than those between the 4-MO and 6-MO calculation, and 
results from the fact that electronic correlations between the active orbitals are smaller 
than those between the active and frozen orbitals. However, the 2-MO basis is clearly not 
capable of predicting accurate energies for the key excitations in biphenyl. The errors 
from the 6-MO calculation for the $1^3B_{1u}$, $1^1B_{1u}$ and $3^1A_{g}$ states 
are 20\%, 13\% and 12\%, respectively. 

Before turning to a discussion of the 2-MO calculation for oligophenylenes, let us use the 
biphenyl results to predict the main features of the optical excitations in oligophenylenes. 
First, the excitations are either $z$- or $y$- polarised states. The $z$-polarised states are 
derived from the $1^1B^-_{1u}$ and $1^1E^-_{1u}(z)$ parent states of benzene. The 
former of these is dipole forbidden, but in forming delocalised states these excitations 
mix and oscillator strength is transfered from the high energy excitation to the low energy 
excitation. The lowest state becomes the $1^1B^-_{1u}$ exciton, and arises 
predominately from excitations between molecular orbital valence and conduction bands. 
The highest energy excitation becomes the localised (or non-bonding) exciton at ca.\ 6.2 
eV, and results from excitations between the non-bonding bands. In the next section we 
will be attempt to describe these two excitations within a 2-MO model.

The $y$-polarised states are derived from the $1^1B_{2u}^+$ and $1^1E_{1u}^-(y)$ 
excitations of benzene. These excitations do not mix in oligophenylenes, and the former 
state is expected to result in the weakly particle-hole allowed feature at 4.5 eV, while the 
latter results in the stronger absorption at 5.5 eV \cite{rice}. 

\section{Oligophenylenes: The Wannier 2-MO Model}
\label{PPP}

In the absence of electronic correlations, the lowest long axis-polarised, dipole allowed 
state would be a transition between the molecular orbital valence and conduction bands 
depicted in Fig.\ 2. The high energy non-bonding exciton arises from transitions between 
the non-bonding orbitals. Turning on the correlations leads to a mixing of the single 
particle basis which, as shown earlier in this paper, leads to both quantitative and 
qualitative corrections. In this section, however, we assume that the low-lying excitations 
may be described entirely within the 2-MO subspace associated with the valence and 
conduction bands. The localised Wannier MOs, which are used to calculate the 
two-electron parameters, are obtained by Fourier transforming the Bloch functions associated 
with these bands. Such Wannier MOs are not only delocalised over neighbouring 
phenylene rings, but they also contain an admixture of different primitive phenylene 
MOs. They retain, however, the same spatial symmetry as their corresponding primitive 
phenylene MOs. 

\begin{table}[htbp]

\centering

\begin{tabular}{|c|c|c|}
\hline Parameter & Name & Wannier-MO  \\
\hline U & Onsite Coulomb repulsion & 5.687  \\
\hline V & Nearest neighbour Coulomb repulsion & 3.612  \\
\hline X & Onsite exchange & 0.581 \\
\hline P & Onsite pair hop & 0.581  \\
\hline $\Delta^*$ & HOMO-LUMO gap & 5.462  \\
\hline $t_{11}=-t_{22}$ & Nearest neighbour hybridisation & 0.667  \\
\hline $t_{12}=-t_{21}$ & Nearest neighbour hybridisation & 0.0 \\
\hline
\end{tabular}

\caption{The parameters and their values in eV used in the 2-MO model (Eq.\ 8) for 
oligophenylenes. $*$ Includes static Coulomb terms from the 
frozen electrons.} 

\label{parameters}
\end{table}

The procedure for obtaining these Wannier orbitals is described in Appendix A. The
two-electron parameters retained in the Hamiltonian and their values are shown in 
table~\ref{parameters}, along with the renormalised one-electron integrals. Notice that 
the three and four centre Coulomb terms are neglected for the oligomer calculation as 
they have little effect on the results. The interactions which will go into the model are: the 
HOMO-LUMO gap, direct onsite and nearest neighbour MO Coulomb repulsion, 
spin-exchange
 and pair hop between MOs on the same repeat unit,
 and hopping between neighbouring 
repeat units. These parameters are the minimum required to model the formation and 
delocalisation of singlet and triplet excitons along the poly({\em p}-phenylene) 
backbone. The 2-MO model Hamiltonian is thus,
\begin{eqnarray}
H & = & -\sum_{i\,\alpha\,\beta\,\sigma} t_{\alpha\beta} \left[ 
a_{i\alpha\sigma}^{\dagger} a_{i+1\beta\sigma} + \rm{ h.c.}\right] + \sum_{i\,\alpha} 
\epsilon_{\alpha}(n_{i\alpha}-1) + U \sum_{i\,\alpha} \left(n_{i\alpha\uparrow}-
\frac{1}{2}\right) \left(n_{i\alpha\downarrow}- \frac{1}{2}\right)
\nonumber \\
& &
+\;\frac{U}{2} \sum_{i\,\alpha\neq\beta} (n_{i\alpha}-1) (n_{i\beta}-1)
+ V \sum_{i \,\alpha\,\beta} (n_{i\alpha}-1) (n_{i+1\beta}-1)
\nonumber \\
& &
-\;X \sum_{i\,\alpha\neq\beta}\left[ \bm{S}_{i\alpha}.\bm{S}_{i\beta}
+\frac{1}{4}\left(n_{i\alpha}-1\right)
 \left(n_{i\beta}-1\right) \right]
\nonumber\\
& &
+\;\frac{P}{2} \sum_{i\,\alpha\neq\beta\,\sigma} a_{i\alpha\sigma}^{\dagger} 
a_{i\alpha\bar{\sigma}}^{\dagger} a_{i\beta\bar{\sigma}} a_{i\beta\sigma},
\label{model},
\end{eqnarray}
where,
$\bm{S}_{i\alpha}=\sum_{\rho \rho^\prime}a_{i\alpha\rho}^{\dagger}
\bm{\sigma}_{\rho\rho^\prime}a_{i\alpha\rho^\prime}$
and $\bm{\sigma}$ are the Pauli spin matrices.

Eq.\ \ref{model} is solved exactly for oligomers of up to six repeat units using the 
conjugate gradient method. For longer oligomers of up to 15 units we use the density 
matrix renormalisation group method \cite{white,gehring}. The DMRG is a powerful, robust, 
portable and highly accurate truncated basis scheme for the solution of low dimensional 
quantum lattice systems, and is especially well suited to the solution of open linear chains 
such as \Ref{model}. Since the method is discussed at length in \cite{white} and reviewed in \cite{gehring} we restrict ourselves here to a discussion of the specifics of our implementation for \Ref{model}.

In addition to the total charge
$\hat{N} = \sum_{i\,\alpha}n_{i\alpha}$
and the total $z$-spin 
$\hat{S}^{z}\tsub{T} = \frac{1}{2} \sum_{i\,\alpha} \left(n_{i\alpha\uparrow} - 
n_{i\alpha\downarrow}\right)$,
the spatial inversion
($\hat{C}_{2}$: $a_{i\alpha\sigma} \mapsto a_{L-i+1\,\alpha\sigma}$),
particle-hole
($\hat{J}$: $a_{i1\sigma}^{\dagger} \mapsto {\rm sgn}(\sigma)a_{i2\bar{\sigma}}$, 
$a_{i2\sigma}^{\dagger} \mapsto {\rm sgn}(\sigma)a_{i1\bar{\sigma}}$)
and spin flip
($\hat{P}$: $a_{i\alpha\sigma} \mapsto a_{i\alpha{\bar\sigma}}$)
symmetry operators are used as good quantum numbers in diagonalising the Hamiltonian. We verify the validity of the DMRG solution by checking that the results obtained for the 
trimer and the pentamer agree with exact diagonalisation results. Basis truncation occurs 
for larger chains. We retain $m=210$ states per block in our calculations. We test the 
convergence of the truncation scheme by examining the non-interacting ($U = V = X = P 
= 0$) case which can easily be diagonalised exactly for any chain length. We have found that the DMRG resolves gaps between these states well and truly above the accuracy required in order to make comparisons with experiments, that is, a few hundreds of an eV. The accuracy is even better in the interacting case where states are more localised and gaps are widened \cite{white}. A systematic analysis of the convergence of energies in \Ref{model} is 
presented elsewhere \cite{lavrentiev}.

Table~\ref{ppp1} shows the energies of the vertical transitions as a function of oligomer 
length of the $1^3B_{1u}^+$, the $1^1B_{1u}^-$ and the lowest even, {\em covalent} 
excitation, which we label as $m^1A^+_g$. Notice that, in general, the $m^1A^+_g$ is 
{\em not} the lowest $A_g$ excitation. For example, in biphenyl, it corresponds to the 
$3^1A^+_g$ state; there being an {\em ionic} $A_g$ state below
 it. Furthermore, it is not necessarily the state with the largest oscillator strength with the $1^1B_{1u}^-$ state. Also shown are the experimental results for crystalline thin films. 
Evidently, for longer oligomers, for which the Wannier parameterisation should be the 
most appropriate, the theoretical predictions are ca.\ 1 eV too high.  Although these 
results are not unreasonable, we nonetheless conclude that a 2-MO model, 
whose parameters are obtained directly form the underlying P-P-P model, is incapable of 
quantitative predictions of the low-lying exciton energies.

\begin{table}[htbp]

\centering

\begin{tabular}{|c|c|c|c|c|}
\hline $L$ & $1^1B^-_{u}$ & $m^1A^+_g$ & $1^3B^+_u$ & Experimental Optical 
Gap \\
\hline 2 & 5.13 & 7.02 & 4.27 & 4.80(a) \\
\hline 3 & 4.79 & 6.79 & 3.99 & 4.5(b) \\
\hline 4 & 4.66 & 6.66 & 3.88 & --- \\
\hline 5 & 4.59 & 6.61 & 3.83 & --- \\
\hline 6 & 4.56 & 6.58 & 3.80 & 3.9(b) \\
\hline 7 & 4.52 & 6.56 & 3.78 & --- \\
\hline 11 & 4.51 & 6.54 & 3.76 & --- \\
\hline 13 & 4.50 & 6.54 & 3.75 & --- \\
\hline 15 & 4.50 & 6.54 & 3.75 & --- \\
\hline $\infty$ & 4.50 & 6.54 & 3.75 & 3.43(b), 3.3(c), 3.5(d) \\
\hline
\end{tabular}

\caption{Calculated vertical transition energies in eV for oligophenylenes of various 
lengths, $L$, using the Wannier MO parameters.  Note that the $m^1A^+_g$ exciton 
is the {\em lowest covalent} $A_g$ singlet excited state. Experimental results from 
biphenyl crystals (a) \citex{crystal} and crystalline films (b) \citex{shacklette}, (c) 
\citex{tieke}, (d) \citex{ambrosch}.} 

\label{ppp1}
\end{table}

In the 2-MO model excitations between the valence and conduction bands and between 
the non-bonding orbitals are decoupled. As described above, the former leads to the 
$1^1B_{1u}$ exciton, while the latter results in the non-bonding exciton. The energy of 
this exciton can therefore be estimated: it is the energy of the singlet benzene exciton 
predicted by the 2-MO model in section~\ref{benzene}. This is $6.43$ eV, which is quite 
close to the experimental value of ca.\ $6.2$ eV. 

\section{Conclusions}
\label{conclusions}

In this paper we have developed a recently introduced two state (2-MO) model for the 
low-lying, long axis-polarised excitations of poly({\em p}-phenylene) oligomers and 
polymers. We have shown that a 2-MO model, based on the HOMO and LUMO states, 
can be derived from the underlying Pariser-Parr-Pople (P-P-P) model by freezing out 
those orbitals further from the Fermi energy. By a careful comparison of the predictions 
of the 4-MO and 2-MO models to the exact P-P-P results for benzene and biphenyl we 
have shown quantitatively how the 2-MO model fails to predict excitation energies. For 
example, it predicts the $1^1B_{1u}$ exciton to be 0.66 eV (13\%) higher than the exact 
P-P-P calculation in biphenyl. 

Next, we have solved the 2-MO model, where the MOs are Wannier orbitals obtained by 
Fourier transforming the Bloch states associated with the valence and conduction bands 
of poly({\em p}-phenylene) polymers, for oligophenylenes of up to 15 repeat units 
using the DMRG method. We showed that these parameters lead to an over estimation of 
the $1^1B_{1u}$ exciton by ca.\ $1$ eV in comparison with experiment. 

Thus, both a comparison of the 2-MO model to full P-P-P calculations and to experiment 
shows that it quantitatively fails to predict the excitation energies. The reason for these 
discrepencies lies in the fact that the original HOMO and LUMO single particle basis 
does not provide an adequate representation for the many body processes of the electronic 
system. Thus, the orbitals which are assumed to be frozen in the 2-MO model in fact 
participate dynamically in the many body states. To incorporate this effect, one can 
assume that a two state model, with the relevant many body interactions, is appropriate to 
describe the low energy physics, but that these parameters are renormalised from their 
bare P-P-P values.  The aim is to parameterise a two state model by fitting its predictions 
to the exact P-P-P model calculations of benzene and biphenyl. If a robust 
parameterisation of oligophenylenes can be achieved, by which we mean a reasonably 
accurate prediction of exciton energies, then other quantities, such as oscillator strengths, 
non-linear optical coefficients and correlation functions can be calculated with some 
confidence. This is
the subject of \cite{lavrentiev}.

R.J.B.\ acknowledges the support of the Australian Research Council. The DMRG 
calculations were performed on the SGI Power Challenge facility at the New South 
Wales Centre for Parallel Computing. W.B.\ acknowledges financial support from the 
EPSRC (U.K.) (GR/K86343), and thanks Dr. M. Yu.\ Lavrentiev for useful discussions.  
H.D.\ is supported by an EPSRC studentship.

\appendix

\setcounter{equation}{0}
\def\theequation{\thesection.\arabic{equation}}

\section{Wannier Orbitals}

The Bloch states associated with the band $j$ with energy $\epsilon_k^j$ are
\begin{equation}
\diracm{\psi_k^j}=\sum_{\alpha} u_k^{j\alpha} \diracm{\psi_k^{\alpha}}
\end{equation}
where $\sum_{\alpha} u_k^{j\alpha}$ is the eigenvector. 

$\diracm{\psi_k^{\alpha}}$ is the Fourier transform of the $\alpha$ phenylene molecular 
orbital state, i.e.
\begin{equation}
\diracm{\psi_k^{\alpha}} = \frac{1}{\sqrt{N}} \sum_{n=1}^N 
\diracm{\psi_n^{\alpha}} e^{ikn}.
\end{equation}
Likewise, the Wannier function associated with the $j$th Bloch state is,
\begin{equation}
\diracm{\psi_n^j} = \frac{1}{\sqrt{N}} \sum_{n=1}^N \diracm{\psi_k^j} e^{-ikn}.
\end{equation}

Substituting (A1) and (A2) into (A3) gives
\begin{equation}
\diracm{\psi_n^j} = \sum_m \sum_{\alpha} \tilde{u}_m^{j\alpha} 
\diracm{\psi_{n+m}^{\alpha}},
\end{equation}
where $ \tilde{u}_m^{j\alpha}$ is the Fourier transform of $ u_k^{j\alpha}$, i.e.
\begin{equation}
\tilde{u}_m^{j\alpha} = \frac{1}{N} \sum_k u_k^{j\alpha}e^{ikm}.
\end{equation}
The one-electron integrals are easily obtained within the Wannier MO basis. We require
\begin{equation}
t_{mn}^{j^{\prime}j} = \matrixelm{ \psi_m^{j^{\prime}} } { {\cal H}_{1-e} } { 
\psi_n^{j} }.
\end{equation}

Now, using (A3)
\begin{equation}
{\cal H}_{1-e}\diracm{\psi_n^{j}}= \frac{1}{\sqrt{N}} \sum_k \epsilon_k^j 
\diracm{\psi_k^j} e^{-ikn}
\end{equation}
and by virtue of the orthogonality of the Bloch states, i.e.
\begin{equation}
\innerm{ \psi_{k^{\prime}}^{j^{\prime}} }{ \psi_k^{j} } = 
\delta_{k^{\prime}k}\delta_{j^{\prime}j}
\end{equation}
\begin{equation}
t_{mn}^{j^{\prime}j} = \frac{1}{N} \delta_{j^{\prime}j} \sum_k \epsilon_k^j e^{ik(m- 
n)}.
\end{equation}

The two-electron integrals are more easily calculated by using the real space 
representation and inserting the Ohno interaction, as described in section~\ref{benzene}.


\begin{thebibliography}{99}
\bibitem{footnote}
We work in a real space representation (and thus construct Wannier MOs), and not a 
$k$-space representation (and thus use Bloch states), because the DMRG method is more 
readily implemented in real space.
\bibitem{footnote2}
It is interesting to note that although electron-electron interactions lead to a significant 
difference in the energies from the non-interacting case, there is a rather small effect on 
the wavefunctions. Thus, in the non-interacting limit ($U=0$) all four singlet and triplet 
states are degenerate with an energy $2t\tsub{p}=5.078$ eV. The wavefunctions are 
$\diracm{1^1B_{1u}} = f_1(\diracm{S_{24}} - \diracm{S_{35}})$, 
$\diracm{1^1E_{1u}(z)} = f_2(\diracm{S_{24}} + \diracm{S_{35}})$, 
$\diracm{1^1B_{2u}} = f_3(\diracm{S_{25}} - \diracm{S_{34}})$, 
$\diracm{1^1E_{1u}(y)} = f_4(\diracm{S_{25}} + \diracm{S_{34}})$, 
$\diracm{1^3B_{1u}} = f_5(\diracm{T_{24}} - \diracm{T_{35}})$, 
$\diracm{1^3E_{1u}(z)} = f_6(\diracm{T_{24}} + \diracm{T_{35}})$, 
$\diracm{1^3B_{2u}} = f_7(\diracm{T_{25}} - \diracm{T_{34}})$ and 
$\diracm{1^3E_{1u}(y)} = f_8(\diracm{T_{25}} + \diracm{T_{34}})$, where 
$f_i=1/\sqrt{2}$, $\diracm{S_{ij}} = \frac{1}{\sqrt{2}} \sum_{\sigma} 
a_{i\sigma}a_{j\sigma}^{\dagger}\diracm{GS}$, $\diracm{T_{ij}} =   
-a_{i\downarrow}a_{j\uparrow}^{\dagger}\diracm{GS}$ and \dirac{GS} is the ground 
state wavefunction. Although turning on the interactions leads to an admixture of higher 
energy states, the wavefunctions are still dominated by the `single-particle' excitations. 
The coefficients are now $f_1=0.696,$ $f_2=0.681$, $f_3=0.672$, $f_4=0.681$, 
$f_5=0.680$, $f_6=0.683$, $f_7=0.703$ and $f_8=0.683$.
\end{thebibliography}
\end{document}